\documentclass[11pt]{article}
\usepackage{amsfonts}
\usepackage{amssymb,latexsym}
\usepackage{amsmath,amscd}
\usepackage{theorem}

\theorembodyfont{\rm}

\numberwithin{}{}

\input{tcilatex}

\begin{document}

\title{\textbf{Some geometric features of Berry%
\'{}%
s phase}}
\author{Alejandro Cabrera\thanks{\texttt{cabrera@mate.unlp.edu.ar}} \\
Departmento de Matem\'{a}tica, Universidad Nacional de La Plata\\ Argentina\\
}
\maketitle

\begin{abstract}
In this letter, we elaborate on the identification and construction of the
differential geometric elements underlying Berry%
\'{}%
s phase. Berry bundles are built generally from the physical data of the
quantum system under study. We apply this construction to typical and
recently investigated systems presenting Berry%
\'{}%
s phase to explore their geometric features.
\end{abstract}

Berry%
\'{}%
s phase \cite{Berry} discovery for parameter-dependent quantum systems
showed the existence of fundamental differential geometric features in
quantum physics. In fact, the \emph{geometric nature} of this phase leads to
both, its theoretical importance and the ability to perform experiments in
which this phase can be detected \cite{ShWil}. Because of this fact, it is
important to have a suitable description of the underlying physical data of
the system in terms of the geometric objects which lead to the phase shift
under study.

Usually \cite{nonabel} for \emph{non abelian} phases, this geometric setting
is modeled by means of the universal $U(k)$ bundle over the grassmannian
manifold $G_{K^{m}}(\mathcal{H})$ \cite{Hus} of $K^{m}-$dimensional
subspaces of the total Hilbert space $\mathcal{H}$, endowed with its
canonical connection. When the parameter $b$ varies within a parameter space
$\tilde{B}$, the $K^{m}-$dimensional \textit{eigenspace} $F_{b}^{m}\subset
\mathcal{H}$ of a given \textit{energy} $\epsilon ^{m}(b)$ also varies
describing a curve in the grassmannian. Parallel transport along this curve
captures the geometric Berry phase effect.

The aim of this letter is to enlighten the fact that the geometry directly relevant
for the study of the underlying physical problem is not that of the above mentioned
universal bundle, but the one of the \emph{pull back} \cite{Hus} bundle along the
induced map Parameter Space$\longrightarrow $Grassmannian. Indeed, the space of
\emph{physical parameters} can be much smaller than or have a very different
geometry from that of the grassmannian manifold. We
remark this in the same sense that the geometry of a $2-$sphere $%
S^{2}\hookrightarrow \mathbb{R}^{3}$, even though following form that of $%
\mathbb{R}^{3}$, is different and can be independently studied from that of the
bigger ambient space $\mathbb{R}^{3}$. A very simple example of this is given by a
large ($s\geq 1$) spin $s$ in an external magnetic field. In that case, the dimension of the
grassmanninan can be huge (equal to $4s$) while the space of physically accessible eigenspaces $%
F_{b}^{m}\subset $ $\mathcal{H}$ through the manipulation of the external magnetic
field, is a submanifold of at most dimension $3$ for every $s$ (see examples below).

On the other hand, considering the pull back bundle has the strategical
advantage of allowing for a direct study on how the geometry of the physical
parameter space $B$ affects the resulting Berry phase. Moreover, this pull
back bundle $U(E^{m})\longrightarrow B$, that we shall refer to as \emph{%
Berry bundle}, can be directly constructed from the \emph{natural physical inputs
}defining the quantum system: the Hilbert space of states $\mathcal{H} $ and the
parameter dependent Hamiltonian $H(b)$; with no further reference to the universal
bundle. The construction itself yields the relevant topology of the \emph{allowed
parameter space} $B\subset \tilde{B}\ $(\cite{import par spac}). In the next
sections, we give the details of this basic and direct construction of the Berry
bundle, which captures the essential geometric features of the parameter dependence
of the system. In this context, the universal bundle has an \emph{a posteriori
}appearance due only to its universality property \cite{Hus}.

Finally, we remark that this construction has also the advantage of \emph{%
computability}: in most cases, one can concretely build and characterize the Berry
bundle together with the corresponding \emph{Berry connection} on it. The reason is
that, typically, the allowed parameter space $B$, which is the base of the Berry
bundle, is just $\mathbb{R}^{n}$ minus \emph{singular points}. Consequently, a
direct characterization of the bundle in terms of an open cover and transition
functions becomes doable.

The reader might also find interesting the fact that the above mentioned basic
ingredients of bundle theory, v.g. transition functions and local connection
expressions, seem to be hand-tailored for the sole study of Berry phases.

At the end of this letter, as suggested in \cite{Levay}, we apply the mentioned construction to obtain new results
on the global geometry underlying typical, and recently investigated, concrete
quantum\emph{\ }systems presenting Berry phase effects.

\textbf{Setting and notation.}-\ \label{subsec:setting}\label%
{subsubsec:GPintro}We now recall some well known facts about (non abelian)\
Berry%
\'{}%
s phases. Let \ $(\mathcal{H},H(b))$ be a \textit{Quantum system}, where the
\textit{Hamiltonian} operator $H(b):\mathcal{H}\longrightarrow \mathcal{H}$
depends smoothly on \textit{parameters} $b\in \tilde{B}$, for $\tilde{B}$ a
manifold from where, a priori, the parameters can be chosen.

For each parameter $b,$ $P^{m}(b):\mathcal{H}\longrightarrow F_{b}^{m}$
denotes the corresponding \textit{orthogonal projection}. As usual, the
\textit{evolution operator}\textbf{\ }$U_{t,t_{0}}:\mathcal{H}%
\longrightarrow \mathcal{H}$ is a unitary operator s.t. $\psi
(t)=U_{t,t_{0}}\psi (t_{0})$, for $\psi (t)$ denoting the state of the
system at time $t$. Consider $b(t):I=[t_{i,}t_{f}]\longrightarrow \tilde{B}$
a piecewise smooth curve on the \textit{parameter space}\textbf{\ }$\tilde{B}
$. If the time evolution of the parameters $b(t)$ is \emph{slow}, we can
assume that this evolution is \emph{adiabatic}, and thus it is a good
approximation to assume that if $\psi (t_{0})\in F_{b(t_{0})}^{m}$ then $%
\psi (t)\in F_{b(t)}^{m}$ for each $t\in I$. Consequently, the evolution
operator can be written as\cite{Non adiabat} $U_{t,t_{0}}=\underset{m}{%
\Sigma }U_{t,t_{0}}^{m}$, where the blocks are linear (unitary) maps $%
U_{t,t_{0}}^{m}=P^{m}(b(t))U_{t,t_{0}}P^{m}(b(t_{0})):F_{b(t_{0})}^{m}%
\longrightarrow F_{b(t)}^{m}$.

Consider, as usual, a curve $R_{t,t_{0}}:I\longrightarrow U(\mathcal{H})$ of
unitary linear operators in $\mathcal{H}$ taking the eigenspace at time $%
t_{0}$,$\ F_{b(t_{0})}^{m}$, to the eigenspace at time $t$, $F_{b(t)}^{m}$.
If we write $U_{t,t_{0}}^{m}=R_{t,t_{0}}\tilde{U}%
_{t,t_{0}}^{m}=R_{t,t_{0}}P^{m}(b(t_{0}))\tilde{U}_{t,t_{0}}^{m}$ where $%
\tilde{U}_{t,t_{0}}^{m}\in U(F_{b(t_{0})}^{m})$ is a linear unitary
endomorphism of the subspace $F_{b(t_{0})}^{m}\in \mathcal{H}$, it follows
from Schrodinger equation that $\tilde{U}%
_{t,t_{0}}^{m}=U_{dyn}^{m}(t)U_{geom}^{m}(t)$, the product of the \emph{%
dynamical }and \emph{geometrical} part of the solution. The (non abelian)\
geometric part is defined by%
\begin{equation}
\frac{d}{dt}U_{geom}^{m}(t)U_{geom}^{m\dagger
}(t)=-P^{m}(b(t_{0}))R_{t,t_{0}}^{-1}\frac{d}{dt}R_{t,t_{0}}P^{m}(b(t_{0})).
\label{eq:GPinfinitesimal}
\end{equation}%
When the parameter space curve $b(t)$ is closed , i.e. $b(t_{0})=b(t_{f}),$
then the unitary operator $U_{geom}^{m}(t_{f})$ is the (non-abelian)
geometric \textbf{Berry phase} (factor) for the underlying adiabatic
parameter-dependent quantum system. In what follows, the geometric nature of
$U_{geom}(t_{f})$ shall become clear.

\textbf{Constructing the Berry bundles over parameter space}.-\label%
{subsec:Vector Bundle E} We shall now construct a vector bundle $E^{m}$ over
the \emph{allowed parameter space} $B\subset \tilde{B}$, capturing the
essential geometry underlying the parameter dependence of the system. This
construction can be straightforwardly adapted to the case of adiabatic and
invariant operators as suggested in \cite{invar op}. This bundle must have
as fiber over each parameter $b\in B$, the vector space of all possible
eigenstates of $H(b)$ with eigenvalue $\epsilon ^{m}(b)$. For, under an
\emph{adiabatic} change of parameters $b(t),$ the system evolution will be
described by a curve $\psi (t)$ in $E^{m}$ projecting onto $b(t)$ in the
base $B$. To begin the construction, we assume some smoothness conditions on
the parameter dependence of $H$:

$\left( i\right) $ we can smoothly (in $b$) choose the eingenvalues $%
\epsilon ^{m}(b)$ of $H(b)$, $m=1,2,...$ via smooth maps $B\longrightarrow
\mathbb{R}$\ taking $b\longmapsto \epsilon ^{m}(b)$ such that $%
det(H(b)-\epsilon ^{m}(b)id_{\mathcal{H}})=0$ for all $b\in B$.

$\left( ii\right) $ the degeneracy degree $K_{m}:=dim(F_{b}^{m})$ of the
energy level $\epsilon ^{m}(b)$ is constant for all $b\in B$.\bigskip

The space of \textit{allowed parameters} $B$ is just $\tilde{B}$ minus
singular points where the above conditions do not hold. Removing singular
points creates holes yielding a non trivial topology on $B$ and allowing for
non trivial bundles (underlying Berry%
\'{}%
s phase)\ over it\cite{import par spac}. Note that this parameter space
topology arises naturally within our construction from the physical inputs.

\textbf{Vector Berry bundle} $E^{m}\longrightarrow B$.- Fix an energy label $%
m$ and consider the map $\mathfrak{F}^{m}:B\times \mathcal{H\longrightarrow }%
B\times \mathcal{H}$ as $\mathfrak{F}^{m}(b,\alpha ):=(b,H(b)\alpha
-\epsilon ^{m}(b)\alpha )$. This is a vector bundle morphism and, since the
dimension of the eigenspaces of $H(b)$ is assumed to be constant, the rank
of $\mathfrak{F}$ is the same on all fibers. So $E^{m}:=Ker(\mathfrak{F}%
^{m}) $ is also a vector bundle over $B$ which we shall refer to as the
\emph{vector Berry bundle}. As mentioned in the introduction, this vector
bundle can be also obtained from a canonical one by pull back via the map $%
B\longrightarrow G_{K^{m}}(\mathcal{H})$. Note that, as desired, $E^{m}=Ker(%
\mathfrak{F}^{m})=\underset{b\in B}{\sqcup }F_{b}^{m}$ and the projection is
given by $E^{m}\overset{\pi }{\longrightarrow }B$, $\left\vert
m(b),i\right\rangle \longmapsto b$ when $\left\vert m(b),i\right\rangle \in
F_{b}^{m}$, $i=1...K_{m}$. Moreover, the vector bundle $E^{m}$ is endowed
with the \textbf{fiber metric }induced by that of the Hilbert space $%
\mathcal{H}$. We stress that the geometry of this bundle is determined
directly by the physical data involving the dependence of the Hamiltonian $%
H(b)$ on the parameters $b$.

\textbf{The principal }$U\left( K_{m}\right) -$\textbf{Berry bundle }$%
U(E^{m})$\textbf{\ over }$B$\textbf{.}- \label{sec:U(E)}We now perform the
analogue of passing to describe the system with the evolution operator in $U(%
\mathcal{H})$ instead of using time dependent states in $\mathcal{H}$.
Moreover, the construction we give below can be applied to any Hermitian\
vector bundle $E^{m}$ over a smooth manifold $B$, showing that the relevant
geometric data of the problem is already encoded in $E^{m}$. Consider the $%
U\left( K_{m}\right) $-principal bundle of \emph{orthonormal bases\ }$%
\{\left\vert m(b),i\right\rangle \}$ of $E^{m}$; $\pi
:U(E^{m})\longrightarrow B$, $\{\left\vert m(b),i\right\rangle \}\mapsto b$
on which the Lie group $U(K_{m})$ acts on the \emph{right} via $\{\left\vert
m(b),k\right\rangle \}\cdot (a_{j}^{i})=\{\left\vert \widetilde{m(b),k}%
\right\rangle =\Sigma _{j}\ a_{j}^{k}\left\vert m(b),j\right\rangle \}$. We
shall refer to $U(E^{m})\longrightarrow B$ as the \emph{Berry Bundle} over
the allowed parameter space $B$. As stated in the introduction, this
principal bundle $U(E^{m})$ can be obtained from the corresponding universal one by \emph{%
pulling back} via the function $B\longrightarrow G_{K^{m}}(\mathcal{H)}$.
Choosing a $b_{0}\in B$ and using the bijection $U(E^{m})\equiv
\dbigsqcup\limits_{b}\{linear\ maps\ \hat{u}:F_{b_{0}}^{m}\longrightarrow
F_{b}^{m}\ such\ that\ \left\langle \hat{u}v,\hat{u}w\right\rangle _{%
\mathcal{H}}=\left\langle v,w\right\rangle _{\mathcal{H}}\}$, every element $%
\{\left\vert m(b),k\right\rangle \}\in U(E^{m})$ can be seen as a map $\hat{u%
}:F_{b_{0}}^{m}\longrightarrow F_{b}^{m}$.

\textbf{Local and global geometries}.- It can be seen\cite{Hus} that, for
every $b_{0}\in B$, there exists an \emph{open patch} $U_{b_{0}}\subseteq B$
of $b_{0}$ and a smooth map
\begin{equation}
R:U_{b_{0}}\longrightarrow Isom(F_{b_{0}}^{m}\longrightarrow \mathcal{H})
\label{Eq: R}
\end{equation}%
such that, for each $b\in U_{b_{0}},\ \{R_{b}\left\vert
m(b_{0}),i\right\rangle \}_{i=1}^{K_{m}}$ is a (\emph{moving})\emph{\
orthonormal basis\ }of the eigenspace above parameter $b$, $%
F_{b}^{m}\subseteq \mathcal{H}$. Above $\{\left\vert m(b_{0}),i\right\rangle
\}_{i=1}^{K_{m}}$ is a fixed orthonormal basis of $F_{b_{0}}^{m}$ and $%
Isom(F_{b_{0}}^{m}\longrightarrow \mathcal{H})$ denotes the manifold of
linear isometries from $F$ into $\mathcal{H}$. The map $R_{b}$ of eq. $%
\left( \ref{Eq: R}\right) $ defines smooth \textbf{local} \textbf{sections}
of the Berry bundles $E^{m}$ and $U\left( E^{m}\right) $ by $\sigma
_{i}^{R}(b)=R_{b}\left\vert m(b_{0}),i\right\rangle $ and $\Sigma
^{R}(b)=R_{b}P^{m}(b_{0}):F_{b_{0}}^{m}\longrightarrow F_{b}^{m}$,
respectively. Note, however, that the bundles might not admit \emph{global
sections} because, in the intersection of patches, different moving basis of
$F_{b}^{m}$ might be glued together in a nontrivial fashion. This \emph{%
global geometry}, fixed by the $b$ dependence of the system%
\'{}%
s Hamiltonian $H(b),$ is captured by the Berry bundle%
\'{}%
s the \emph{transition functions}\cite{Kobay} $\psi _{\alpha \beta
}:U_{\alpha }\cap U_{\beta }\subset B\longrightarrow U(K_{m})\simeq
U(F_{b_{0}}^{m})$ given by$\ $%
\begin{equation}
\psi _{\alpha \beta }(\ b)=\dsum\limits_{i,j=1}^{K_{m}}\left\langle w_{\beta
}^{j}(b)|w_{\alpha }^{i}(b)\right\rangle \left( \left\vert
w^{j}(b_{0})\right\rangle \left\langle w^{i}(b_{0})\right\vert \right)
\label{Eq: trans funct}
\end{equation}%
where $\left\vert w_{\alpha ,\beta }^{i}(b)\right\rangle =R_{b}^{\alpha
,\beta }\left\vert w^{i}(b_{0})\right\rangle $ for $R_{b}^{\alpha ,\beta }:$
$U_{\alpha ,\beta }\longrightarrow U(E^{m})$ local sections on two
intersecting patches $U_{\alpha }$ and $U_{\beta }$ and $\left\{ \left\vert
w^{i}(b_{0})\right\rangle \right\} $ a fixed o.n. basis for the fiber $%
F_{b_{0}}^{m}$ over a chosen $b_{0}\in B$.

\textbf{Berry connection giving the geometric phase}.-\label{subsec:conect
on U(E)} From eq. $\left( \ref{Eq: R}\right) $, take $%
R_{t,t_{0}}P^{m}(b_{0})\ $as $R_{b(t)}\in U(E^{m})$ and $U_{geom}^{m}(t)\in
U\left( F_{b_{0}}^{m}\right) \simeq U\left( \mathbb{C}^{K_{m}}\right) $ to
be determined. It can be easily seen that there exists a globally defined
\textbf{principal connection}\cite{Kobay} $A^{m}:TU(E^{m})\longrightarrow
u(K_{m})=Lie(U(K_{m}))$, that we shall call the \textbf{Berry connection},
on the principal fiber bundle $U(E^{m})\overset{\pi }{\longrightarrow }B$
such that its \emph{local expression} along a section $\Sigma ^{R}$ is%
\begin{equation*}
(\Sigma ^{R\ast }A^{m})_{b}=P^{m}(b_{0})\ \left. R^{-1}(b)\right\vert
_{F_{b}^{m}}d_{b}R\ P^{m}(b_{0})
\end{equation*}%
\begin{equation}
=\dsum\limits_{k=1}^{dimB}\dsum\limits_{i,j=1}^{K_{m}}\left\langle w^{j}(b)|%
\frac{\partial }{\partial b_{k}}|w^{i}(b)\right\rangle \left( \left\vert
w^{j}(b_{0})\right\rangle \left\langle w^{i}(b_{0})\right\vert \right) \
db_{k},  \label{Eq: A local express}
\end{equation}%
where where $b_{k}$ are local coordinates on $U_{b_{0}}\subset B$.

Then, eq. $\left( \ref{eq:GPinfinitesimal}\right) $ for the geometric phase
is equivalent to requiring the curve $R_{t,t_{0}}P^{m}(b_{0})\cdot
U_{geom}^{m}(t)$ in the Berry bundle $U(E^{m})$ to be \emph{horizontal }with
respect to the Berry connection. Consequently, $R_{t,t_{0}}P^{m}(b_{0})\cdot
U_{geom}^{m}(t)$ is the \emph{parallel transport} of the initial condition $%
id_{\mathcal{H}}P^{m}(b_{0})\cdot U_{geom}^{m}(t_{0})$ along $b(t)\in B$.
When the parameter curve is closed $b(t_{0})=b(t_{f})$ and $%
U_{geom}^{m}(t_{0})=id_{F_{b_{0}}^{m}}$, parallel transport from patch to
patch yields a global and geometrically defined \emph{holonomy}\cite{Kobay} $%
U_{geom}^{m}(t_{f})\in U\left( F_{b_{0}}^{m}\right) $ which is precisely the
(non abelian if $K_{m}>1$) Berry phase\cite{Berry} factor associated to the
underlying quantum system.\bigskip

\textbf{Geometry of Berry bundles for}\emph{\ }$B\simeq S^{2,1}$.-\label%
{subsec: B spheres} The present approach allows for the use of the physical
data encoded in the topology of the parameter space $B$ of the system under
study. Indeed, assuming that $B$ is (smooth) homotopically equivalent to the
sphere $S^{k=1,2}$, permits to yield conclussions on the geometry of the
Berry bundle for a general $H(b)$, by means of some bundle-theoretic results.

$\left( I\right) $ For $k=2$ it holds that if $E^{m}\longrightarrow B$ is
orientable as a vector bundle, then the principal $U(K_{m})$-bundle $%
U(E^{m})\longrightarrow B$ is trivializable.

$\left( II\right) $ For $k=1$, then $U(E^{m})\longrightarrow B$ is always
trivializable.

It is clear that, when the Berry bundle is trivializable, local
considerations extend to the whole parameter space $B$ by the existence of
smooth \emph{global} sections. This simplifies the analysis and shows that
there is no \emph{nontrivial geometric contribution }to the GP.

\textbf{Spin in magnetic field}.- \label{ex:spin in mag field}Below, we
elaborate on the global geometry underlying the example presented by Berry
\cite{Berry}. Let $\mathcal{H}=V_{spin(s)}$ be the Hilbert space
corresponding to a particle with spin $s$. For $b\in \tilde{B}:=\mathbb{R}%
^{3}$, let $H(b)=g\ $%
h{\hskip-.2em}\llap{\protect\rule[1.1ex]{.325em}{.1ex}}{\hskip.2em}
$b\cdot \hat{s}$ be the Hamiltonian giving the interaction between this
particle and a magnetic field represented by $b$. For a fixed $b\in \mathbb{R%
}^{3},$ the $(2s+1)\ $eingenvalues of $H(b)$ are $\epsilon ^{m}(b)=g\ $%
h{\hskip-.2em}\llap{\protect\rule[1.1ex]{.325em}{.1ex}}{\hskip.2em}
$\left\Vert b\right\Vert m$, all with degeneracy $1$ except for $b=\mathbf{0}
$ where all eigenvalues collapse to only one with full degeneracy. We thus
see that the largest submanifold $B\subset \mathbb{R}^{3}$ satisfying
condition $(ii)$ above is $B:=\mathbb{R}^{3}-\{\mathbf{0}\}$ which,
topologically, is equivalent to a $2-$sphere. Which is the Berry bundle for
this spin systems? Below we answer this question.

Fix $m\ $s.t. $-s\leq m\leq $ $s$. Since $B\approx \mathbb{R}_{>0}\times
S^{2}$, we can cover $B$ with two open sets $U^{\pm }:=\mathbb{R}_{>0}\times
\left( S^{2}-\{b_{0}^{\pm }\}\right) $ over which the Berry bundles are
trivial. Indeed, if $b_{0}^{\pm }=(0,0,\pm 1)\in S^{2}$ denote the poles,
each $b\in U^{\pm }$ can be taken by a rotation to the pole. This rotation
induces a spin rotation $\hat{R}^{\pm }(b)$ taking the eigenvector $%
\left\vert b_{0}^{\pm },m\right\rangle \in F_{b_{0}}^{m}$ to $\left\vert
b^{\pm },m\right\rangle \in F_{b}^{m}$ and can be smoothly chosen for each $%
b $ defining local sections on $U^{\pm }$ as in $\left( \ref{Eq: R}\right) $%
\cite{Non bnorm dep}.

The \emph{isomorphism class}\cite{Hus} of the $U(1)$ principal Berry bundle $%
U(E^{m})$ over $B$ for this system is thus determined by the \emph{homotopy
class} of the transition function restricted to the where the two patches
intersect, i.e., by $\psi _{+-}:\mathbb{R}_{>0}\times \{equator\ of\
S^{2}\}\longrightarrow U(1)\simeq \pi ^{-1}(b_{0}^{+})=\{\hat{u}%
:E_{b_{0}^{+}}^{m}\circlearrowleft \}$. A straightforward calculation of the
\emph{transition function }$\left( \ref{Eq: trans funct}\right) $ at $%
b=\left\Vert b\right\Vert (cos\varphi ,sin\varphi ,0)\in \mathbb{R}%
_{>0}\times \{equator\ of\ S^{2}\}$ yields%
\begin{equation*}
\psi _{+-}(b)=e^{i2m\varphi }\left\langle
b_{0}^{+},m|b_{0}^{-},m\right\rangle \ \left\vert b_{0}^{+},m\right\rangle
\left\langle b_{0}^{+},m\right\vert .
\end{equation*}%
Since we can take $\left\vert b_{0}^{\pm },m\right\rangle =\left\vert \pm
m\right\rangle $ with $\hat{S}_{z}\left\vert \pm m\right\rangle =\pm
m\left\vert \pm m\right\rangle $ and, since $\left\langle
b_{0}^{+},m|b_{0}^{-},m\right\rangle $ never vanishes, we conclude that the
transition map $\psi _{+-}$ \emph{winds }the equator of $S^{2}\ 2m-$times in
$U(1)$. This winding number \emph{specifies the above mentioned isomorphism
class of the bundle} $U(E^{m})$. Note that the homotopy class of the
transition map above does not depend on $\left\Vert b\right\Vert $, as
expected. Moreover, this class depends \emph{only }on the \bigskip $z-$axis
spin projection $m$ and \emph{not on the total spin} $s$. From this follows
the fundamental result that, for particles of spin $s$ and $s$%
\'{}
such that $m$ is an allowed value of the $z-$axis spin projection for both,
then $E^{m,s}\simeq E^{m,s%
{\acute{}}%
}$ and $U(E^{m,s})\simeq U(E^{m,s%
{\acute{}}%
})$. For example, any fermion (boson) shall yield the same $U(E^{m=\frac{1}{2%
}})$ ($U(E^{m=0})$) modulo isomorphism. Concretely, for $m=\frac{1}{2}$, the
transition function is characterized by the winding number $1$, and so the
Berry bundle is isomorphic to the well known \emph{Hopf fibration} $%
SU(2)\approx S^{3}\longrightarrow S^{2}$\cite{Hus}. For $m=1$, the winding
number is $2$, and then the Berry bundle is isomorphic to the bundle $%
SO(3)\longrightarrow S^{2}$, $r\longmapsto r\cdot b_{0}^{+}\ $related to
classical \textit{rigid body geometric phases}\cite{Mont}.

From the local expression $\left( \ref{Eq: A local express}\right) $ for the
Berry connection on $U^{+}$, it can be easily seen that the holonomy $%
Hol(\Gamma )\in U(1)\simeq \{\hat{u}:E_{b_{0}^{+}}^{m}\circlearrowleft \}\ $%
of any simple closed path $\Gamma $ in $B$ (measured from the identity of $%
U(1)$)\ reads $Hol(\Gamma )=exp(-im\Omega (\Gamma ))\ \left\vert
b_{0}^{+},m\right\rangle \left\langle b_{0}^{+},m\right\vert $, where $%
\Omega (\Gamma )$ is a (signed)\ solid angle having $\Gamma $ as boundary.
We thus recover the expression first obtained by Berry \cite{Berry}.

\textbf{Non abelian phase from holonomic quantum computation}.- We take the
following example from \cite{Florioetal} in the context of \emph{holonomic
quantum computation}. Non abelian Berry phases play a crucial role in the
theoretical construction of fault tolerant quantum gates because of their
\emph{geometric nature}. The setting is the following: $\mathcal{H}%
=Span\{\left\vert 0\right\rangle ,\left\vert 1\right\rangle ,\left\vert
a\right\rangle ,\left\vert e\right\rangle \}$ and $H(\overrightarrow{\Omega }%
)=\left\vert e\right\rangle \left( \Omega _{0}\left\langle 0\right\vert
+\Omega _{1}\left\langle 1\right\vert +\Omega _{a}\left\langle a\right\vert
\right) +h.c.$ for $\overrightarrow{\Omega }=(\Omega _{0},\Omega _{1},\Omega
_{a})\in \mathbb{R}^{3}$ being the associated Rabbi frequencies.

The eigenvalues are $\epsilon ^{0}(\overrightarrow{\Omega })=0$ and $%
\epsilon ^{\pm }(\overrightarrow{\Omega })=\pm \left\Vert \overrightarrow{%
\Omega }\right\Vert $. We see that for $\overrightarrow{\Omega }\neq 0$,
conditions $(i)(ii)$ given above are satisfied and that $dimE_{%
\overrightarrow{\Omega }}^{0}=2=const$. So we take $B=\mathbb{R}^{3}-\{0\}$
as in the previous example. Note that in this case, since $dimE_{%
\overrightarrow{\Omega }}^{0}=2$, the structure group of the associated
Berry bundle $U(E^{0})\longrightarrow B$ is $U(2)$ and, thus, the induced
Berry phase shall be \emph{non abelian}.

Identifying the $\mathbb{R}^{3}$ vector components $\check{1}\equiv \check{x}%
,\ \check{0}\equiv \check{y},\ \check{a}\equiv \check{z}$, it is easy to
define local sections of the bundle $E^{0}\longrightarrow B$ in $U^{\pm }$
as in the previous example. Also as before, we calculate the transition
function $\psi _{+-}:\mathbb{R}_{>0}\times \{equator\ of\
S^{2}\}\longrightarrow U(2)\simeq \pi ^{-1}(\vec{\Omega}_{0}^{+})=\{\hat{u}%
:E_{\vec{\Omega}_{0}^{+}}^{0}\circlearrowleft \}$ which determines the
global geometry of the Berry bundle. For$\ \overrightarrow{\Omega }%
=\left\Vert \overrightarrow{\Omega }\right\Vert \cdot (sin\alpha ,cos\alpha
,0)\in \mathbb{R}_{>0}\times \{equator\ of\ S^{2}\}$\cite{Ej q hol comp} $\ $%
\begin{equation*}
\psi _{+-}(\alpha )\equiv \left(
\begin{array}{cc}
-cos2\alpha  & -sin2\alpha  \\
-sin2\alpha  & cos2\alpha
\end{array}%
\right) \in U(2)
\end{equation*}%
for $\alpha \in S^{1}$ parameterizing the equator of $S^{2}$. The above $U(2)
$ matrix is the product of a constant one by an $\alpha $-dependent matrix
staying within $SU(2)\subset U(2)$ for all $\alpha $. From this, it follows
that the vector Berry bundle $E^{0}\longrightarrow B$ is orientable. Thus,
by result $\left( I\right) $ given above, the $U(2)\ $bundle $%
U(E^{0})\longrightarrow B$ is \emph{isomorphic to the trivial bundle }$%
B\times U(2)$. In fact, once this is known, it is not hard to find global
sections which are different from the local ones presented in reference \cite%
{Florioetal}, where the analysis carried out is thus not global. This result
may contribute to clarify some aspects of the existence of the fidelity
revivals discussed in \cite{Florioetal}.

\textbf{Topological phases}.- When the Berry connection $A^{m}$ $\left( \ref%
{Eq: A local express}\right) $ on the Berry principal fiber bundle $U(E^{m})%
\overset{\pi }{\longrightarrow }B$ is \emph{flat}\cite{Kobay} $dA-[A,A]=0$,
then Berry%
\'{}%
s phase becomes \textbf{topological}. This means that the phase or holonomy
which the final state gains depends only on the homotopy class\ of the
parameter curve $b(t)\in B$ and not on its geometry any more . When the
structure group is abelian, the flatness condition reads $dA=0$\cite{Flat R3}%
. Topological phases appear in various distinct areas\cite{cites top}, below
we give two simple examples.

\textbf{Spin on the plane and quantum Hall effect in graphene.}- For $J=1,2$%
, consider the Hamiltonian $H_{J}(b)=\varepsilon \ \hat{s}\cdot \check{n}%
_{b}^{J}$ for $\hat{s}$ being the vector of $\hat{S}_{i}$ matrices
corresponding to spin $s$, $\check{n}_{b}^{J}=-(cosJ\varphi _{b},sinJ\varphi
_{b})$, with $\varphi _{b}$ the angle giving the direction of the external
parameter $b$ on the $xy$ plane $\mathbb{R}^{2}$ and $\varepsilon $ a
constant. This example with $J=1$ follows straight forwardly from our first
example (as done in \cite{Berry}), but we study it independently bellow to
illustrate our general construction of the bundle $U(E^{m})$.

In this case, the largest submanifold $B\subset \mathbb{R}^{2}$ satisfying
the given conditions $(i)(ii)$ is $B=\mathbb{R}^{2}-\{\mathbf{0}\}$ since at
$b=\mathbf{0}$, all eigenvalues collapse to only one with full degeneracy.
From result $(II)$ above, we know that the bundle $U(E^{m})$ is trivializable%
\cite{Section spin on plane}. The eigenstates are nondegenerated, so the
Berry phase is abelian and the Berry connection $\left( \ref{Eq: A local
express}\right) $ is
\begin{equation*}
A^{m}(b)=i\ J\ s\ d\varphi _{b}
\end{equation*}%
for which, clearly, $dA^{m}=0$. Consequently, the Berry connection is \emph{%
flat} and the associated Berry phase is \emph{topological}. Explicitly, for
a path $b(t)\ $in $B$ the phase reads%
\begin{equation}
U_{geom}^{m}(t_{f})=exp\left( -i2\pi s\ WN(b(t))\right)  \notag
\end{equation}%
where $WN(b(t))$ denotes the winding number of $b(t)$ around $(0,0)$ in the $%
xy-$plane.

The case $J=1$, describes a spin $s$ interacting with a magnetic field $b$
varying on the plane. So, when $WN(b(t))=+1$, for \emph{fermions} (half integer $s$%
)\ the above phase factor is $-1$ whereas for \emph{bosons }(integer $s$) it
is $+1$ independently of the $\hat{S}_{i}$ eigenvalue $m$. This reproduces
the results of \cite{Berry}.

Finally, monolayer and bilayer \textit{graphene Berry phases} correspond to $%
J=1,2$, respectively\cite{Graph}. For this systems, the parameter $b$
represents the carrier%
\'{}%
s momentum within $2d$ graphene. When a magnetic field is present, carriers
describe closed trajectories in the plane, so the parameter $b(t)$ describes
a simple closed loop surrounding $(0,0)\ $in $B$ after one carrier
revolution. Then, the associated topological Berry phase factor is%
\begin{equation}
U_{geom}^{J,m}(t_{f})=exp\left( -i2J\pi s\right)   \notag
\end{equation}%
which, for $s=\frac{1}{2}$, is $(-1)^{J}$ as explained in \cite{Graph}.

\noindent \textbf{Acknowledgements: }A.C. wants to thank Prof. J. Solomin
for stimulating discussions and suggestions, and to CONICET\ Argentina for
financial support.


\begin{thebibliography}{99}
\bibitem{import par spac} Berry \cite{Berry} first noted the crucial role
that singularities in parameter space played in the resulting geometric
phase.
\bibitem{Non bnorm dep} This sections does not depend on $\left\Vert
b\right\Vert $ as expected.
\bibitem{Ej q hol comp} This is the matrix of the transition function in a
chosen basis $\left\vert w_{+}^{j}(\vec{\Omega}_{0}^{+})\right\rangle $ of $%
E_{\vec{\Omega}_{0}^{+}}^{0}$, with $\vec{\Omega}_{0}^{+}$ the north pole of
$S^{2}$.
\bibitem{Flat R3} Moreover, for $U(1)$ bundles and $B\subset \mathbb{R}^{3}$
then the flatness condition takes the more familiar form $\nabla \times \vec{%
A}=0$.
\bibitem{Section spin on plane} Indeed, for a fixed $m$, the section $%
\left\vert m(b)\right\rangle =exp(i\varphi _{b}s)exp\left( i\varphi _{b}\hat{%
S}_{z}\right) \left\vert b_{0},m_{y}\right\rangle $ is globally defined in $%
B $.
\end{thebibliography}
\end{document}